\documentstyle[preprint,epsf,aps,prb]{revtex}

\begin{document}

\newlength{\figwidth}
\setlength{\figwidth}{\textwidth}
\addtolength{\figwidth}{-.2\figwidth}

\tighten
\draft 

\title{Shot noise in resonant tunneling structures}
\author{G. Iannaccone\cite{email}, M. Macucci, and B. Pellegrini}
\address{
Dipartimento di Ingegneria dell'Informazione: 
Elettronica, Informatica e Telecomunicazioni \\
Universit\`a degli studi di Pisa, Via Diotisalvi 2, 
I-56126 Pisa, Italy}
\date{Submitted to Phys. Rev. B}
\maketitle

\begin{abstract} 
We propose a quantum mechanical approach to noise in 
resonant tunneling structures,
that can be applied
in the whole range of transport regimes,
from completely coherent to completely incoherent. In both
limiting cases, well known results which have
appeared in the literature are recovered.
Shot noise reduction due to both Pauli exclusion and Coulomb repulsion,
and their combined effect, are studied as a function of the rate of
incoherent processes in the well (which are taken into account
by means of a phenomenological relaxation time), and of temperature.
Our approach allows the study of noise in a variety of operating
conditions (i.e., equilibrium, sub-peak voltages, 
second resonance voltages), and as a function of temperature,
explaining experimental results and predicting interesting new results,
such as the dependence of noise on filled emitter states and the
prediction of both increasing and decreasing shot noise with 
increasing temperature, depending on the structure.
It also allows the determination of the major contributions to shot 
noise suppression by performing noise measurements at the second
resonance voltage.
\end{abstract}


\narrowtext
\section{Introduction}

In recent years, researchers have shown increasing interest 
in noise in resonant tunneling structures. In fact, from an
experimental point of view, noise measurements can provide
information about the structure and the transport properties of
resonant tunneling devices complementary to those given by DC 
characteristics and small signal AC responses. On the other
hand, the correct prediction of noise properties is a good
check for the validity of transport models for such devices.

In 1989, Lesovik \cite{lesovik89} predicted that, in 
the case of completely coherent transport, shot noise 
could be lower than its classical value for totally 
independent electron crossings through the structure,
i.e., the so called full shot noise;\cite{vanderziel86} 
in 1990, Li and coworkers \cite{li90} showed the first 
experimental evidence of such phenomenon in double barrier
diodes.
Since then, many theoretical studies appeared in the literature, 
based on both coherent 
\cite{yurkkoch90,buettiker90,chenting91,buettiker92} 
and semiclassical models, 
\cite{chenting92,davihyld92,eguehers94,brown92,chen93,hungwu93,birkdejo95,hersdavi93} 
while few experimental results are available.
\cite{li90,brown92,birkdejo95,ciambrone94,liu95,ciambrone95}
If the time of observation $T$ is much longer than the average time $\tau_T$ an
electron takes to traverse the double barrier, the noise spectral density 
at low frequencies ($\ll 1/\tau_T$) can be reduced only if consecutive current
pulses are correlated, i.e., if pulse distribution is 
sub-poissonian.\cite{vanderziel86}
Two are the mechanisms which have been considered responsible for
introducing such correlation:
Pauli exclusion,
\cite{chenting92,davihyld92,eguehers94}
and electrostatic repulsion, \cite{brown92,chen93}
which both tend to smooth fluctuations of the number of electrons in the well region.

In agreement to what seems to be confirmed by experimental measurements,
most theoretical studies predict a maximum suppression of
one half of the classical shot noise value, that can be obtained if the 
transmission probabilities of the two barriers are equal.\cite{spiegare}
However, most astonishing appears the fact that such results
have been obtained both with 
coherent\cite{lesovik89,yurkkoch90,buettiker90,chenting91,buettiker92,davihyld92,liu95}
and semiclassical\cite{chenting92,davihyld92}
models, and even if time correlations between consecutive traversals
of the two barriers are discarded.\cite{macupell93}

In this paper, we propose an approach 
addressing noise properties of generic resonant tunneling
structures in the whole range of transport regimes, from 
completely coherent to completely incoherent, from a 
quantum mechanical viewpoint.
In the limit of coherent
transport the result of Lesovik \cite{lesovik89} is recovered,
while, in the opposite limit of loss of coherence for all
electrons traversing the well, the semiclassical
results of Davies \cite{davihyld92} are obtained.
We also consider the combined effects of Pauli exclusion and
of Coulomb repulsion, and show that maximum shot noise
suppression of one half is to be expected independently of
the coherence of transport, at least up to a given amount of
collisions in the well.

We also study noise behaviour of resonant tunneling 
structures in various bias conditions, i.e.,
equilibrium, sub-peak voltages, and second resonance voltages.
In particular, as a check for our model, we recover the
Johnson-Nyquist\cite{johnson27} noise at equilibrium. Moreover,
we study noise dependence on temperature, which has
been measured experimentally,\cite{ciambrone95} but has
received little attention from a theoretical point of view.

The outline of the paper is as follows: in Sec.\ II we discuss our
model for transport in resonant tunneling structures and for transitions
through the barriers, and introduce the simplifications and
approximations needed to address the problem analytically. In Sec.\ III
we calculate time-dependent currents and the current power spectral
density, while in Sec.\ IV we obtain noise in typical operating conditions,
i.e., various applied voltages and operating temperatures. The particular
case of large well structures (where the
characteristic time for fluctuations of the number of carriers
in the well is larger than the time of observation) is addressed in
Sec.\ IV. The Conclusion section ends the paper.

\section{Model}

In a recent paper,\cite{ianna_unified95} it has been shown that the 
sequential tunneling approach can be adopted to describe the whole
range of transport regimes in resonant tunneling structures,
from purely coherent to completely incoherent.
This approach is applicable when the structure can
be seen as consisting of three isolated regions $\Omega_l$,
$\Omega_w$, and $\Omega_r$, i.e.,
the left reservoir, the well region, and the right reservoir,
respectively, 
that are weakly coupled through the two tunneling barriers 1 and
2, as is sketched in Fig. 1, so that 
first order perturbation theory is applicable.

Let each allowed state in $\Omega_l$, $\Omega_w$, $\Omega_r$ be
characterized by a set of parameters
$\alpha_l$, $\alpha_w$, $\alpha_r$, respectively. The density of
states and the occupation factor in region $\Omega_s$ ($s=l,w,r$)
are $\rho_s(\alpha_s)$ and $f_s(\alpha_s)$, respectively.
Following Bardeen, \cite{bardeen61} tunneling is simply
treated as an electronic transition between levels
in different regions. Given that
first order perturbation theory is applicable, tunneling 
probabilities per unit time are given by the 
Fermi ``golden rule.''\cite{sakurai}

We also take into account the effects of elastic and inelastic collisions
in the well by means of a single
phenomenological relaxation time $\tau_{\rm coll}$: an electron
in the well has a probability $dt/\tau_{\rm coll}$ of experiencing
a collision in the infinitesimal time interval $dt$, and electrons 
emerge from collisions with a thermal quasi-equilibrium energy
distribution and a completely random phase. 
Based on a similar model (in which a relaxation length was used
instead of a relaxation time), a compact formula for the density of states in
a quantum well has been obtained.\cite{ianna_compact95}.

It is worth noticing that in the relaxation time approximation\cite{aschmerm77} 
all collisions are effective in randomizing phase and relaxing energy.
For simplicity, we adopt this model, and we do not discuss the
details of energy dependence of the relaxation time. In fact,
for the purpose of this paper, we just need to be aware of the
fact that collisions affect the density of states by broadening
and lowering the resonance peaks,\cite{stonelee85,jonsgrin87,buettiker88}
and affect also the occupation
factor $f_w$ in the well, which 
we divide into three components,\cite{ianna_unified95}
\begin{equation}
f_w(\alpha_w) = f_{w0}(\alpha_w) + f_w^l(\alpha_w)
+ f_w^r(\alpha_w)
,\label{fw}
\end{equation}
where $f_{w0}$ is the Fermi-Dirac occupation probability 
associated to the quasi-Fermi level $E_{fw}$ in the well, 
and $f_w^l$ and $f_w^r$ are the occupation factors for
electrons which have come from the left and the right
electrode, respectively, and have not undergone a collision in the well.

Suppose that $N$ electrons are in the well: the probability that in the
time interval $dt$ an electron enters the well through barrier $m$
($m=1,2$) is $g_m(N)dt$. Following Davies {\em et al.} \cite{davihyld92}
we call $g_m(N)$ ``partial generation rate'' for barrier $m$. The 
probability that in the time interval $dt$ an electron escapes from
the well through barrier $m$ is $r_m(N)dt$, where $r_m(N)$ is 
the ``partial recombination rate'' for barrier $m$. Of course, we can
also define the total generation rate $g(N) \equiv g_1(N) + g_2(N)$,
and the total relaxation rate $r(N) \equiv r_1(N) + r_2(N)$.

\subsection{Generation and recombination rates}

Let $\hat{H}_m$ ($m=1,2$) be the perturbation Hamiltonian due
to barrier $m$.
We start by considering tunneling through the first barrier:
following Bardeen,\cite{bardeen61} we obtain the matrix element
for an electron transition from a state $|\alpha_l \rangle$
in $\Omega_l$ to a state $|\alpha_w \rangle$ in $\Omega_w$. We indicate it
with $M_{1lw} \equiv \langle \alpha_w | H_1 | \alpha_l \rangle$.
The probability per unit time $\nu_{(l \rightarrow |\alpha_w \rangle )}$
that an electron in $\Omega_l$ 
jumps into an unoccupied
 state $| \alpha_w \rangle$ in $\Omega_w$ is given by the Fermi
``golden rule'' (to first order in $\hat{H}_1$):\cite{sakurai}
\begin{equation}
\nu_{(l \rightarrow |\alpha_w \rangle )} =
\frac{2 \pi}{\hbar} \int |M_{1lw}|^2 \rho_l f_l
d\alpha_l
\label{nulw}
.\end{equation}
Therefore, the transition rate $g_1$ from $\Omega_l$ to $\Omega_w$ is 
obtained by integrating $\nu_{(l \rightarrow |\alpha_w \rangle )}$ over all unoccupied
states in $\Omega_w $, i.e.,
\begin{eqnarray}
g_1 & = & \int \nu_{(l \rightarrow |\alpha_w \rangle )} \rho_w (1 - f_w) d\alpha_w \nonumber \\
    & = & \frac{2 \pi}{\hbar} \int \int |M_{1lw}|^2 \rho_l
		\rho_w f_l (1 - f_w) d\alpha_l d\alpha_w
\label{g1iniz}
.\end{eqnarray}

Analogously, we can obtain the recombination rate $r_1$: the probability
per unit time $\nu_{(|\alpha_w \rangle \rightarrow l)}$ that an electron in a state 
$| \alpha_w \rangle$ in $\Omega_w$ jumps in $\Omega_l$ is
\begin{equation}
\nu_{(|\alpha_w \rangle \rightarrow l)} \equiv \frac{2\pi}{\hbar}
	\int |M_{1lw}|^2 \rho_l (1 - f_l) d\alpha_l        
\label{nuwl}
,\end{equation}
where we have used $M_{1lw} = M_{1wl}^*$;
now $r_1$ is easily obtained by integrating 
$\nu_{(|\alpha_w \rangle \rightarrow l)}$ over occupied states in $\Omega_w$, i.e.,
\begin{eqnarray}
r_1 & = & \int \nu_{(|\alpha_w \rangle \rightarrow l)} \rho_w f_w d\alpha_w \nonumber \\
    & = & \frac{2\pi}{\hbar} \int \int |M_{1lw}|^2 \rho_l \rho_w f_w (1-f_l)
			d \alpha_l d \alpha_w
.\label{r1iniz}
\end{eqnarray} 

We wish to point out that both $g_1$ and $r_1$ 
are functionals of $f_w$,
which appears explicitly in (\ref{g1iniz}) and
(\ref{r1iniz}), and, through the Poisson equation,
affects the total Hamiltonian of the system, and hence both the densities 
of states $\rho_w$, $\rho_l$, and the transition matrix elements
$M_{1lw}$. Since we are interested in considering the
effects on shot noise of Pauli exclusion and Coulomb repulsion 
associated to electrons occupying allowed states in
the well, it should also be noticed that the 
effect of Pauli exclusion is accounted for in (\ref{g1iniz})
through the term $(1-f_w)$, while Coulomb 
repulsion affects $g_1$ and $r_1$ through the term $|M_{1lw}|^2$, which 
depends on the potential energy profile modified by the charge 
accumulated in the well.

The passages from (\ref{nulw}) to ({\ref{r1iniz}) and the above considerations  
can be repeated for transitions through barrier 2, by simply 
substituing pedices $1$ with $2$, and $l$ with $r$.

\subsection{Simplifying assumptions}

In order to study the effects on shot noise of fluctuations in 
the distribution of the occupied states in the well, we have to make some 
simplifying, but justified, assumptions.

First, we assume that the occupation factors $f_l$ and $f_r$
for states in the external regions do not fluctuate, which means that 
thermalization mechanisms in these regions are highly 
effective in establishing an equilibrium distribution.
Then, we also assume that the 
effect of fluctations of $f_w$ 
on the potential profile is weak enough that the densities of states
in the three regions can be considered as constant.
Therefore, fluctuating terms in $g_1$ and $r_1$
are $f_w$, of course, and $|M_{1lw}|^2$, which
depends on $f_w$ via the Poisson equation.

In realistic structures, the well region contains many states,
which definitely makes not tractable the problem of considering
$g_1$ and $r_1$ as functionals of
the occupation factor of each level in the well. 
Hence, we need to make a further strong assumption:
that generation and recombination
rates depend on $f_w$ only through the total
number of electrons in the well $N$, defined as 
\begin{equation}
N \equiv \int \rho_w f_w d\alpha_w 
\label{enne}
.\end{equation}
In other words, we assume
\begin{equation}
\begin{array}{c}
g_m = g_m(N) \\
r_m = r_m(N) 
\end{array} \hspace{1cm} \mbox{for } m=1,2
.\label{grm}
\end{equation}
The total generation
and recombination rates are
\begin{eqnarray}
g(N) & = & g_1(N)+g_2(N) \label{gn} \\
r(N) & = & r_1(N)+r_2(N) \label{rn}
.\end{eqnarray}
The eigenvalues of $N$ are positive integers, and cannot be
greater than the total number of states in the well
$N_0 \equiv \int \rho_w d\alpha_w$.
Though it is not necessary nor rigorous, we will often 
consider $N$ as a number
large enough to be treated as a continuous quantity,
and, for example, will write derivatives of functions
of $N$.

On one hand, it is very reasonable to state that the 
self-consistent energy profile, and therefore $|M_{1lw}|^2$,
depend essentially on the total number of electrons in the
well, and only to a second order on the detailed shape of the
probability density distribution, i.e., on 
which particular states are actually occupied.
On the other hand, we have to note that our assumption
discards the effect of the detailed shape of the
term $f_w$ in both (\ref{g1iniz})
and (\ref{r1iniz}).
We shall consider possible drawbacks of this last
approximation later on, when discussing the results
obtained.

It is worth noticing that the 
expressions for generation and recombination rates used
by Davies {\em et al.}, i.e., $g(N) = (N_0-N)/\tau_e$, and
$r(N)=N/\tau_c$ (Eq. (3.10) of Ref. \onlinecite{davihyld92}), 
are a particular case of (\ref{gn}) and (\ref{rn}). 

\subsection{Steady state distribution of electrons in the well}

The steady state distribution $p_0(N)$ of electrons in the
well can be obtained by using the conditions that $r(0)=0$ and
$N$ has to be not negative along with the detailed balance on the
rates:\cite{davihyld92}
\begin{equation}
r(N+1) p_0(N+1) = g(N) p_0(N) 
,\label{detailed}
\end{equation}
which, by induction, yields
\begin{equation}
p_0(N) = p_0(0)
    \prod_{m=1}^{N} \frac{g(m-1)}{r(m)}
\label{p0nonlin}
.\end{equation}
One can then obtain $p_0(0)$ after imposing probability normalization, i.e.,
\begin{equation}
\sum_{N=0}^{N_0} p_0(N) = 1
.\end{equation}

\subsection{First order approximation of generation and recombination
rates}

If the distribution of the total number of electrons in the well
is narrow enough we can greatly improve the tractability of the 
problem by linearising recombination and generation rates. Let
$\tilde{N}$ be the number for which $g(\tilde{N}) = r(\tilde{N})$
and let us define $\Delta N \equiv N - \tilde{N}$.
We develop transition rates to first order in $\Delta N$, i.e.,
\begin{equation}
g(N) = \left\{ 
	\begin{array}{ll}
	g(\tilde{N}) - \Delta N / \tau_{g} & 
			\mbox{for } \Delta N < g(\tilde{N}) \tau_g \\
	0	& \mbox{for } \Delta N \geq g(\tilde{N}) \tau_g 
	\end{array}
	\right.
		\label{glin}
,\end{equation}
\begin{equation}
r(N) = \left\{ 	
	\begin{array}{ll}
	0 & \mbox{for } \Delta N < - r(\tilde{N}) \tau_r \\ 
	r(\tilde{N})  + \Delta N / \tau_{r} 
		& \mbox{for } \Delta N \geq - r(\tilde{N}) \tau_r
	\end{array}
	\right.
,\label{rlin}
\end{equation}
where the characteristic
times $\tau_{g}$ and $\tau_{r}$ for generation and recombination
are defined as
\begin{eqnarray}
\frac{1}{\tau_{g}} & \equiv & - \left. \frac{dg}{dN}
                        \right|_{N = \tilde{N}} \label{taug} \\
\frac{1}{\tau_{r}} & \equiv & \left. \frac{dr}{dN} 
                        \right|_{N = \tilde{N}} \label{taur}
.\end{eqnarray}
In Fig. 2 a qualitative picture of the approximation made is shown.
Now, by substituting (\ref{glin}) and (\ref{rlin})
into (\ref{p0nonlin}), we obtain
\begin{equation}
\frac{p_0(N + 1)}{p_0(N)} =
\frac{ g({\tilde{N}}) - \Delta N/\tau_g }
	{ r({\tilde{N}}) + (\Delta N + 1)/\tau_r } =
\frac{\tau_r}{\tau_g} \frac{\tilde{N}_0 - L}{L+1}
\label{p0linqua}
\end{equation}
if we define $L \equiv \Delta N + g(\tilde{N}) \tau_r$ and
$\tilde{N}_0 \equiv g(\tilde{N}) (\tau_r + \tau_g)$.
Since, according to (\ref{glin}) and (\ref{rlin}) (Fig. 2),
the number of electrons in the well cannot be lower than 
$\tilde{N} - r(\tilde{N}) \tau_r$, for which the recombination rate
is zero, nor greater than $\tilde{N} + g(\tilde{N}) \tau_g$, for which
the generation rate is zero, then
$p_0(N)$ is non zero only between these limits, and $L$ runs from $0$ to
$\tilde{N}_0$.
From (\ref{p0linqua}) we obtain for $p_0(N)$ 
the binomial distribution
\begin{equation}
p_0(N) = \left( \begin{array}{c} \tilde{N}_0 \\ L \end{array} \right)
                \frac{\tau^{\tilde{N}_0}}{\tau_g^L \tau_r^{\tilde{N}_0 - L}}
\label{statprob}
,\end{equation}
where $\tau^{-1} \equiv \tau_g^{-1} + \tau_r^{-1}$. 
As can be seen, (\ref{statprob}) reduces to Eq. (3.13) of 
Ref. \onlinecite{davihyld92}, if one assumes the transition
rates given by Eq. (3.10) of the same reference,
so that, as a consequence, $\tilde{N}_0$ reduces
to $N_0$, and $L$ to $N$.

We now make the reasonable hypothesis that all the stochastic processes
we are considering are ergodic, and indicate with $ \langle a \rangle$
both the expectation value and the time average of any quantity $a$.
Eq. (\ref{statprob}) allows us
to calculate the average value of $N$ and the variance:
\begin{eqnarray}
\langle N \rangle & \equiv & \sum N p_0(N) = \tilde{N} \label{avn} \\
\mbox{var}(N) & \equiv &\sum N^2 p_0(N) - \langle N \rangle^2
		= \tilde{N}_0 \frac{\tau^2}{\tau_g \tau_r} 
		=\langle g \rangle \tau \label{varn}
.\end{eqnarray}
It is worth noticing that, according to (\ref{varn}), the
standard deviation $[\mbox{var}(N)]^{1/2}$ is much lower than the 
allowed range $\tilde{N}_0$ of variation for $N$, 
therefore we can reasonably assume
the linearization of transition rates to be applicable. Moreover,
given that $\langle N \rangle = \tilde{N}$, we have
\begin{equation}
\langle r \rangle = \langle g \rangle = g(\tilde{N}) = g (\langle N \rangle) 
.\end{equation}

It is useful to linearize partial recombination rates through each barrier,
analogously to what we have done in (\ref{glin}) and (\ref{rlin}) 
for $g(N)$ and $r(N)$. We have to define
$1/\tau_{rm} \equiv (dr_m/dN)|_{N=\tilde{N}}$ and
$1/\tau_{gm} \equiv (-dg_m/dN)|_{N=\tilde{N}}$, $ m=1,2$.
Therefore we have $\langle g_m \rangle = g_m(\tilde{N})$
and $\langle r_m \rangle = r_m(\tilde{N})$, for $m=1,2$. 
The steady-state
current $\langle i \rangle$ at the device electrodes is given by the
net transition rate through either barriers
\begin{equation}
{\langle i \rangle} = q\langle g_1 - r_1 \rangle
= q \langle r_2 - g_2 \rangle
\label{i}
.\end{equation}

We also define quantities that will be used in the following paragraphs:
\begin{equation}
\tau_1^{-1} \equiv \tau_{g1}^{-1} + \tau_{r1}^{-1}, \hspace{0.5cm}  
\tau_2^{-1} \equiv \tau_{g2}^{-1} + \tau_{r2}^{-1}. 
\label{tau12}
\end{equation}
Of course we have
\begin{equation}
\tau_g^{-1}   = \tau_{g2}^{-1} + \tau_{g1}^{-1},  \hspace{0.5cm} 
\tau_r^{-1}  = \tau_{r1}^{-1} + \tau_{r2}^{-1},  
\end{equation}
\begin{equation}
\tau^{-1} = \tau_1^{-1} + \tau_2^{-1} = \tau_g^{-1} + \tau_r^{-1}
\label{tau}
.\end{equation}

\subsection{Autocorrelation function}

The autocorrelation function $c_{NN}(t)$ of $N(t)$ is defined as
\begin{equation}
c_{NN}(t) \equiv \langle \Delta N(0) \Delta N(t) \rangle 
\label{autocordef}
.\end{equation}
By taking the time-derivative we have
\begin{eqnarray}
\frac{d}{dt} c_{NN}(t) &= &
\langle \Delta N(0) \frac{d \Delta N(t)}{dt} \rangle \nonumber \\
& = & \langle \Delta N(0) [ g(N) - r(N) ]\rangle
\label{dcnndt}
,\end{eqnarray}
where we have used the rate equation\cite{davihyld92,vanderziel59} 
$dN(t)/dt = g(N) - r(N)$.
From (\ref{glin}), (\ref{rlin}), (\ref{tau}) and (\ref{autocordef}),
we can write  
$dc_{NN}/dt = - c_{NN}/ \tau$,
from which we finally have
\begin{equation}
c_{NN} = \langle g \rangle \tau e^{-|t|/\tau}
,\label{cnn}
\end{equation}
where we have used (\ref{varn}) and the fact that $c_{NN}(0) = \mbox{var}(N)$.
Let us point out that $\tau$ has the role of characteristic
time of fluctuations in the number of electrons $N$.

\section{Noise}

\subsection{Time-dependent current}

In this section, we are going to calculate the time dependent
current and its noise spectral density in the case of a 
constant voltage applied between the electrodes.
According to the Ramo-Shockley theorem\cite{ramoshoc38}, and to
the electrokinematics theorem \cite{pellegrini86}
that generalises it 
to any system, to the electromagnetic field and to quantum
mechanics, \cite{pellegrini93}
when an electron tunnels through
one of the barriers, it generates a pulse in the current of the external 
circuit \cite{pellmacu93}; the time integral of the current pulse associated to the traversal
of barrier $m$, ($m=1,2$) is $\lambda_m q$, where $\lambda_m$ is equal to
the ratio of the voltage drop across barrier $m$ to the total applied
voltage. Of course, we have $\lambda_1 + \lambda_2 = 1$. In terms
of the quasi-Fermi levels of the three regions, we have
$\lambda_1 = (E_{fl}-E_{fw})/(E_{fl} - E_{fr})$, and
$\lambda_2 = (E_{fw}-E_{fr})/(E_{fl} - E_{fr})$.
Suppose that we observe the system in the interval $(0,T)$,
in which the current $i(t)$ has the form
\begin{eqnarray}
i(t) & = & 
		\lambda_1 q \left[
		\sum_j f_j^{g1}(t-t_j^{g1}) - \sum_j f_j^{r1}(t-t_j^{r1})
		\right] + \nonumber \\
& & 		\lambda_2 q \left[
		\sum_j f_j^{r2}(t-t_j^{r2}) - \sum_j f_j^{g2}(t-t_j^{g2})
		\right]    
,\label{icompleta}
\end{eqnarray}
where $f_j^{g1}$ gives the shape of the current pulse due to a single
generation via barrier 1 starting at time $t_j^{g1}$. 
Traversals of the barrier and current pulse shapes are not 
identical, therefore we
have to associate a different function $f_j^{g1}$ to each pulse.
What all functions $f_j$ have in common is the normalization to unity, and
the fact that their Fourier transform $F_j(\omega)$ is flat and equal
to unity for frequencies much smaller than the inverse of the traversal
time of each barrier. At such frequencies the Fourier transform $I(\omega)$
of the current is \cite{notafour}
\begin{eqnarray}
I(\omega) & \equiv & \int i(t) \exp(-i \omega t) dt \nonumber \\
	& = & 
	\lambda_1 q \left[
	\sum_j \exp(-i\omega t_j^{g1}) - \sum_j \exp(-i\omega t_j^{r1}) \right]
	+ \nonumber \\
& &	\lambda_2 q \left[
        \sum_j \exp(-i\omega t_j^{r2}) - \sum_j \exp(-i\omega t_j^{g2}) \right]
.\label{iomegacompleta}
\end{eqnarray}
The integral and the sums run over all the pulses occuring in the interval of observation
(0,T). We wish to point out that at low frequencies (here ``low''
means again much smaller than the inverse of the traversal time of the
device), the power spectral density of current fluctuations $S(\omega)$
is not influenced by the pulse shape, therefore we do not
expect $S(\omega)$ to be dependent on the 
particular values of $\lambda_1$ and $\lambda_2$ (as 
shown in Appendix B1).

\subsection{Noise spectral density}

The power spectral density $S(\omega)$ of the current is defined as 
\begin{equation}
S(\omega) \equiv \frac{2}{T} \langle |I(\omega)|^2 \rangle - 4 \pi 
		\langle i \rangle^2 \delta(\omega)
\label{sdiomega} 
,\end{equation}
where, again, $T$ is the time of observation.
By substituting (\ref{iomegacompleta}) in (\ref{sdiomega})
we obtain 16 terms of the type
\begin{equation}
\langle 
\sum_k \sum_{j} \exp [ - i \omega (t_j^{\alpha} - t_k^{\beta})]
\rangle
.\label{offdiag}
\end{equation}
where $\alpha,\beta = g_1,g_2,r_1,r_2$.
We can analyze these terms following Ref. \onlinecite{davihyld92}: 
if $\alpha = \beta$ the diagonal terms $j=k$ are equal to
unity, and their sum gives the number of $\alpha$-hops occurring
from time 0 to $T$, i.e., on average, $\langle \beta \rangle T$.

For $\alpha \neq \beta$, or $\alpha=\beta$ and $j \neq k$, we 
can define $h_{\alpha \beta}(t)$
as the probability per
unit time that a $\alpha$-hop occurs at time $t$ given that a
$\beta$-hop occurred at time 0. Therefore we can write (\ref{offdiag})
as
\begin{equation}
\langle
\sum_k \int_0^T \exp [- i \omega(t-t_k^{\beta})] h_{\alpha \beta}(t-t_k^{\beta}) dt
\rangle
\label{offdiag2}
.\end{equation}
As can be seen, the integral in (\ref{offdiag2}) is independent
of time and is simply the Fourier transform $H_{\alpha \beta}(\omega)$
of $h_{\alpha \beta}(t)$. The sum over $k$ in (\ref{offdiag2})
contains, on average, $\langle \beta \rangle T$ terms,
therefore (\ref{offdiag}) becomes
\begin{equation}
\langle \beta \rangle T [ \delta_{\alpha \beta} + H_{\alpha \beta}(\omega)]
\label{sumterm2}
\end{equation}
The detailed derivation of all the correlation functions $H_{\alpha,\beta}(\omega)$
is reported in Appendix A. Substitution of terms like (\ref{sumterm2}) into
$S(\omega)$ finally yields, for $\omega \ll 1/\tau$ (see Appendix B1),
\begin{equation}
\frac{S(\omega)}{2 q^2} = 
\tau^2 \left(  \frac{\langle g_1 + r_1 \rangle}{\tau_2^2} 
	+ \frac{\langle g_2 + r_2 \rangle}{\tau_1^2} \right)
\label{sdiomegabella}
.\end{equation}
As expected, at frequencies smaller enough than the inverse of the
transit time of electrons through the whole device,
the relative sizes of the current pulses corresponding to the traversal
of the two barriers are not relevant, therefore the 
dependence of $S(\omega)$ on $\lambda_1$ and $\lambda_2$ is cancelled out.

We can arbitrarily decompose the net current into a component coming from
the left and going towards the right
$I_l$, given by  the current $q \langle g_1 \rangle $
entering
the well through barrier 1 multiplied by the portion 
$\langle r_2 \rangle /\langle r \rangle$
exiting through barrier 2,
and a component coming from the
right $I_r$, so that we have
\begin{equation}
I_l \equiv q \frac{\langle g_1 \rangle \langle r_2 \rangle}
				{\langle r \rangle},
\hspace{1cm}
I_r \equiv q \frac{\langle g_2 \rangle \langle r_1 \rangle}
				{\langle r \rangle}
\label{il_ir}
.\end{equation}
Substitution of (\ref{il_ir}) into (\ref{sdiomegabella}) yields
(see Appendix C)
\begin{eqnarray}
\frac{S(\omega)}{2q} & = & 
I_l \left[ 1 - \frac{2 \tau^2}{\tau_1 \tau_2} 
		\left( 1 - \frac{\tau_1 \langle r_1 \rangle}
				{\tau_2 \langle r_2 \rangle}
		\right) \right] \nonumber \\
& +&  
I_r \left[ 1 - \frac{2 \tau^2}{\tau_1 \tau_2} 
                \left( 1 - \frac{\tau_2 \langle r_2 \rangle}
                                {\tau_1 \langle r_1 \rangle}
                \right) \right]
\label{sdiomegailir}
.\end{eqnarray}
In the next section we shall apply
(\ref{sdiomegabella}) and (\ref{sdiomegailir})
to derive the expression of noise at several operating conditions.

\section{Noise suppression in different operating conditions}

\subsection{Thermal Equilibrium}

\subsubsection{Conductance at equilibrium}

We shall show that, at equilibrium, 
the noise spectral density $S(\omega)$ given by (\ref{sdiomegailir})
yields the well known Johnson-Nyquist\cite{johnson27} result. 
This is a good test for
our model, and confirms the fact that shot noise and thermal-equilibrium
noise are special forms of a more general noise
formula,\cite{buettiker92,landauer93,stanwilk85,landauer89,sarpdelb93}
i.e., for
the system considered in this paper, Eq. (\ref{sdiomegailir}).

At equilibrium, the occupation factors in each of the three regions
are given by Fermi-Dirac statistics, and, of course, zero net average currents
flow through each barrier, i.e.,
$\langle g_1 \rangle = \langle r_1 \rangle$, and
$\langle g_2 \rangle = \langle r_2 \rangle$.

It is reasonable to suppose that, if we apply an infinitesimal
perturbation from the equilibrium condition,
the electrons in the three regions still
obey quasi-equilibrium distributions, i.e., the occupation factor
for each region has the Fermi-Dirac form, and is associated to a 
quasi-Fermi level which tends to that in equilibrium.

Let us calculate the conductance $G$ at equilibrium.
We need to apply a small voltage $V$ between the right and the left
electrode, as sketched in Fig. 3: we have $qV = E_{fl} - E_{fr}$.
Let us define $\epsilon_1 \equiv E_{fl} - E_{fw}$, and
$\epsilon_2 \equiv E_{fw} - E_{fr}$: so that we also have 
$qV = \epsilon_1 + \epsilon_2$.
The total current is $\langle i \rangle = I_l - I_r $.
From (\ref{i}) we have
\begin{eqnarray}
G & = & \left. \frac{d\langle i \rangle}{dV} \right|_{V=0}\nonumber \\
& = & 
\frac{ q \langle r_2 \rangle }{\langle r \rangle}
\left. \frac{d\langle g_1 - r_1 \rangle}{dV} 
\right|_{V=0} 
+ \frac{ q \langle r_1 \rangle}{\langle r \rangle}
\left. \frac{d\langle r_2- g_2 \rangle}{dV}
\right|_{V=0} 
\label{g_prima}.
\end{eqnarray}

The first derivative can be written in the form
\begin{equation}
\left. \frac{d\langle g_1 - r_1 \rangle)}{d\epsilon_1} 
\right|_{\epsilon_1=0}
\left. \frac{d\epsilon_1}{dV} \right|_{V=0}
= 
\frac{\langle g_1 \rangle}{k_B \Theta}
\left. \frac{d\epsilon_1}{dV} \right|_{V=0}
\protect{\label{bellag1}}
;\end{equation}
where the equality comes from (\ref{utile}), $k_B$ is
the Boltzmann constant and $\Theta$ is the temperature.
By using (\ref{bellag1}) and the 
corresponding equation for transitions through the second barrier,
we write (\ref{g_prima}) as 
\begin{equation}
G = \frac{q}{k_B \Theta \langle r \rangle}
\left(
\langle r_2 \rangle  \langle g_1 \rangle
\left. \frac{d\epsilon_1}{dV} \right|_{V=0} + 
\langle r_1 \rangle \langle g_2 \rangle
\left. \frac{d\epsilon_2}{dV} \right|_{V=0}
\right)
.\label{g_seconda}
\end{equation}
If we remember that at equilibrium $ \langle r_2 \rangle  \langle g_1 \rangle
=  \langle r_1 \rangle \langle g_2 \rangle$, that 
$\epsilon_1 + \epsilon_2 = qV$, and the definition (\ref{il_ir}) of 
$I_l$, we finally obtain $G$ as
\begin{equation}
G = \frac{q I_l}{k_B \Theta}
\label{G}
.\end{equation}

\subsubsection{Thermal Noise}

In order to recover the Johnson-Nyquist noise, we just need to obtain
a simple relation between $\tau_1$ and $\tau_2$ at equilibrium.
From (\ref{taug}), (\ref{taur}), and (\ref{tau}), for the calculation of $\tau_1$ we have to change 
the number of electrons in the well region by a small amount. At equilibrium, 
as we said above, the effect of this operation is to shift the 
quasi-Fermi level in the well with respect to those in the left
and right regions. From (\ref{tau12}) and (\ref{bellag1}) we have
\begin{eqnarray}
\frac{1}{\tau_1} & = &
\left. \frac{d ( r_1 -g_1 ) }{d\epsilon_1} \right|_{
\begin{array}{l} \small \it \epsilon_1 = 0 \\ 
		 \small \it N = \tilde{N} 
\end{array} }
\left. \frac{d\epsilon_1}{dN} \right|_{N = \tilde{N}} \nonumber \\
& = & - \frac{\langle r_1 \rangle}{k_B \Theta} 
\left. \frac{d\epsilon_1}{dN} \right|_{N = \tilde{N}} 
\label{unosutau1}
.\end{eqnarray}
We know that $\epsilon_1 + \epsilon_2 = 0$, because at equilibrium the left and the right
electrode are at the same potential, therefore by comparing
(\ref{unosutau1}) and the corresponding equation for $\tau_2$
we can finally write
\begin{equation}
\tau_1 \langle r_1 \rangle = \tau_2 \langle r_2 \rangle
\label{bellinatau1tau2}
\end{equation}
This is an interesting result: in fact, 
when we put it into (\ref{sdiomegailir}) 
the factors responsible for noise suppression are cancelled out,
and we simply obtain
\begin{equation}
S(\omega) =2 q ( I_l + I_r ) = 4 G k_B \Theta 
\label{thermalnoise}
,\end{equation}
i.e., the Johnson-Nyquist noise
(the second equality comes from (\ref{G}) and the fact that
$ I_l = I_r $ at equilibrium).

\subsection{High bias}

When the voltage $V$ applied between the electrodes is large enough that
most of the electrons injected in the device come from only one
of the electrodes, we say the device is in condition of high bias. 
Without loss of generality, we can assume that positive voltage is applied to the
right electrode, therefore high bias means 
$\langle g_2 \rangle = 0$. The left-going current component 
$I_r$ vanishes too, and the total current 
$\langle i \rangle = I_l $. From (\ref{sdiomegailir})
we can obtain the shot noise factor\cite{vanderziel86} as
\begin{equation}
\gamma \equiv \frac{S(\omega)}{2 q \langle i \rangle} 
= 1 - \frac{2\tau^2}{\tau_1 \tau_2}
\left( 1 - \frac{\tau_1 \langle r_1 \rangle}
		{\tau_2 \langle r_2 \rangle} \right)
\label{gamma}
.\end{equation}

It is worth noticing that, from (\ref{tau}), 
we have
\begin{equation}
\frac{2 \tau^2}{\tau_1 \tau_2} \leq \frac{1}{2}
\label{maximum}
,\end{equation}
and the equal sign holds only if $\tau_1 = \tau_2$. Eq. (\ref{maximum}) implies
that $\gamma \geq 1/2$. We also notice that,
as expected, recombination through barrier one, corresponding to
injected electrons not contributing to the net current, reduces
the shot noise suppression. 
Maximum suppression ($\gamma = 1/2$) 
is obtained only when $\tau_1 =\tau_2$ and $\langle r_1 \rangle$ is zero,
which means that $V$ is high enough
that $\Omega_l$-states in the resonant energy range of 
the well are fully occupied or that electrons in the well occupy states
below the left electrode conduction band edge.

We wish to remember that here $\tau_1$ and $\tau_2$
take into account both the Pauli exclusion principle and space charge effects;
therefore, (\ref{gamma}) means that the combined effects of 
both phenomena cannot push the shot noise factor below $1/2$.

The validity of this conclusion depends on the 
validity of the approximations we have made throughout this paper.
Let us recall the most relevant ones: we have used first
order approximation for generation and recombination rates
as a function of $N$. Actually, this is not very limiting:
in fact strong suppression requires strong correlation
between electron transitions (incorrelated transitions give full
shot noise); and the more transitions are correlated, the less the
number of electrons in the well fluctuates around its average value,
attributing validity to our first order approximation.

The second strong assumption is that generation and recombination rates
depend on the total number of electrons in the well, and not on the  
distribution of occupied levels. This simplification, as we said
above, was necessary to make the problem tractable, but, 
on the other hand, prevents us from evaluating the possible effects 
on noise of ``shape'' fluctuations in the distribution of occupied states in the well.
Inclusion of these effects is possible if one
addresses a very idealized situation, for example a well with only
two allowed levels, where simply two parameters determine the 
electron distribution, as 
Egues {\em et al.}\cite{eguehers94} have done.
In that case, a minimum shot noise factor of approximately $0.45$
is obtained, when the characteristic time for transitions
between the two levels in the well is equal to the characteristic
time for generations from the emitter, and recombinations towards the
collector. 

However, characteristic times for energy relaxation and 
phase randomization in the well in real devices are much lower 
than the characteristic times of escape from
the well, as shown by the poor peak-to-valley ratio 
of experimental devices,
compared to that predicted by completely coherent models. Therefore,
it is very reasonable that at any moment,
for a given number of electrons in the
well, the electron distribution is 
practically the equilibrium one, in a sort of adiabatic approximation.
In such case, our assumption that $g$ and $r$ depend essentially
on $N$ is practically exact: the more inelastic collisions are 
efficient in establishing an equilibrium distribution, the more
we can be confident on $g_m=g_m(N)$, and
$r_m=r_m(N)$, $m=1,2$.

On the basis of these considerations, we do not expect the effects
we have neglected to play a significant role, and the shot noise
factor to drop below one half.
Among available experimental studies, 
only the one by Brown \cite{brown92} exhibits a value of $\gamma$ smaller
than one half, but the estimated accuracy of his results is not reported.

\subsubsection{One dimensional structures}

Most experimental
resonant tunneling structures can be treated
as one-dimensional devices: the problems are therefore
simplified and many quantities of interest 
can be obtained analytically.

In a one dimensional structure, a state in any region can be
decomposed in its longitudinal component $| E\rangle$, its
transverse component $| {\bf k_T} \rangle$, and its spin component
$| \sigma \rangle$, i.e., for $s =l,w,r$,
$ | \alpha_s \rangle = | E_s \rangle \otimes | {\bf k_T} \rangle \otimes
| \sigma \rangle $.
Electron transitions
through either barrier conserve spin, longitudinal energy, and
transverse wave vector, therefore the problem of calculating
generation and recombination rates can be solved just in the longitudinal
direction, and the results can be then integrated over transverse wave
vectors and doubled to account for spin degeneracy (we discard, for
simplicity, other spin effects).
If we discard Coulomb blockade effects, we can use the
one dimensional transition matrix elements derived in
Appendix A of Ref. \onlinecite{ianna_unified95}. For barrier 1
we have 
\begin{eqnarray}
|\langle E_l | H_1 | E_w \rangle|^2 & = &
|M_{1lw}(E_l,E_w)|^2 \nonumber \\
& = &
\hbar^2 \nu_l(E_l) \nu_w(E_w) T_1(E_w) \delta(E_w-E_l)
,\label{m1dim}
\end{eqnarray}
where $\nu_l(E_l)$ and $\nu_w(E_w)$ are usually called attempt frequencies
(because of their resemblance to the classical number of bounces
on the barrier per second)
for the states $| E_l \rangle$ and $| E_w \rangle$, respectively,
and $T_1$ is the tunneling probability of barrier 1.

Let us define $\rho'_s(E_s)$ ($s=l,w,r$) as the density of longitudinal states
in $\Omega_s$, and $\rho_T({\bf k_T})$ the density of transverse states.
From (\ref{g1iniz}) and (\ref{m1dim}), we have
\begin{eqnarray}
g_1 & = & 4\pi\hbar \int dE \int d{\bf k_T} 
	\nu_l(E) \nu_w(E) T_1(E) \rho'_l(E)
                                \rho'_w(E) \times \nonumber \\
& & \rho_T({\bf k_T})
        f_l(E,{\bf k_T}) [ 1 - f_w(E,{\bf k_T}) ] 
\label{g1dim_uno}
.\end{eqnarray}

We assume, for simplicity, that no size effects are present in the 
left electrode, and that the longitudinal density of states $\rho_l$
satisfy the condition $2 \pi \hbar \nu_l(E) \rho'_l(E) = u(E-E_{cbl})$,
where $u$ is the step function, and $E_{cbl}$ is the conduction band
edge of the left electrode. We can write
\begin{eqnarray}
g_1 & = & 2 \nu_w(E_R) T_1^g \times \nonumber \\
& & \int dE u(E-E_{cbl}) \rho'_w 
	\int d{\bf k_T} \rho_T f_l (1-f_w),
\label{g1dim_due}
\end{eqnarray}
where $E_R$ is an arbitrary resonant energy in the well, and 
$T_1^g$ is defined as
\begin{equation}
T_1^g \equiv \frac{\int dE \nu_w T_1 u(E-E_{cbl}) \rho'_w \int 
                d{\bf k_T} \rho_T f_l (1-f_w)}
	{\nu_w(E_R) \int dE u(E-E_{cbl}) \rho'_w \int 
		d{\bf k_T} \rho_T f_l (1-f_w)} 
\label{t1g}
,\end{equation}
i.e., is practically an average of $T_1$ weighted on 
suitable couples of states for transitions from $\Omega_l$ to
$\Omega_w$.
Analogoulsy, for recombination we can write:
\begin{eqnarray}
r_1 & = & 2 \nu_w(E_R) T_1^r \times \nonumber \\
& & \int dE u(E-E_{cbl}) \rho'_w \int d{\bf k_T} \rho_T f_w (1-f_l),
\label{r1dim_uno}
\end{eqnarray}
where $T_1^r$ is defined as
\begin{equation}
T_1^r \equiv \frac{\int dE \nu_w T_1 u(E-E_{cbl}) \rho'_w \int
                d{\bf k_T} \rho_T f_w(1-f_l)}
	{\nu_w(E_R) \int dE u(E-E_{cbl}) \rho'_w \int
                d{\bf k_T} \rho_T f_w(1-f_l)}
.\end{equation}
It is worth noticing that if $\rho'_w$ has a unique strong resonance
for $E=E_R$, then we have $T_1^g = T_1^r = T_1(E_R)$. 

Now, we need to define a few quantities of interest, namely
$N_{l0}$, $\tilde{f}_l$, and $\eta_l$. We have
\begin{eqnarray}
N_{l0} & \equiv & \int dE u(E-E_{cbl}) \rho'_w \int d{\bf k_T} f_l 
,\label{nl0} \\
\tilde{f}_l & \equiv & 
\frac{\int dE u(E-E_{cbl}) \rho'_w \int d{\bf k_T} \rho_T f_l f_w}
     {\int dE u(E-E_{cbl}) \rho'_w \int d{\bf k_T} \rho_T f_w}
,\label{tildefl} \\
\eta_l & \equiv & 
\frac{\int dE u(E-E_{cbl}) \rho'_w \int d{\bf k_T} \rho_T f_w}
     {\int dE \rho'_w \int d{\bf k_T} \rho_T f_w}
\label{etal}
.\end{eqnarray}
$N_{l0}$ has a simple interpretation as the number of electrons with 
longitudinal energies greater then $E_{cbl}$ that would
be in the well if the occupation factor in the well was equal to that
in the left electrode; $\tilde{f}_l$ is the average of $f_l$ over 
occupied states in the well above the conduction band edge of $\Omega_l$,
while $\eta_l$ is the ratio of the number of electrons in the well
with longitudinal energies greater than the conduction band edge of 
$\Omega_w$ to the total number $N$ of electrons in $\Omega_l$
(therefore $0 \leq \eta_l \leq 1$).
From (\ref{nl0}-\ref{etal}), (\ref{g1dim_due}) and (\ref{r1dim_uno}) 
can be written as 
\begin{eqnarray}
g_1 & = & 2 \nu_w(E_R) T_1^g ( N_{l0} - \tilde{f}_l \eta_l N ) 
\label{g1dim_short} \\
r_1 & = & 2 \nu_w(E_R) T_1^r \eta_l N (1 - \tilde{f}_l)
\label{r1dim_short} 
.\end{eqnarray}

To treat the problem analytically,
we assume that $\tilde{f}_l$ and $\eta_l$ are only weakly 
dependent on the number of
electrons in the well, so that the major dependence of $g_1$ and $r_1$
on $N$
are the explicit one in Eqs. (\ref{g1dim_short}) and (\ref{r1dim_short}), 
and those through the 
tunneling probabilities $T_1^g$ and $T_1^r$,
which are due to electrostatic repulsion; the characteristic times
for generation and recombination processes $\tau_{g_1}$ 
and $\tau_{r_1}$ are obtained as
\begin{eqnarray}
\frac{1}{\tau_{g_1}} & = &  - \langle g_1 \rangle
                        \left. \frac{d \ln T_1^g}{dN}
                        \right|_{N = \tilde{N}}
                        + 2 \nu_w(E_R) T_1^g \tilde{f}_l \eta_l
,\label{taug1dim} \\
\frac{1}{\tau_{r_1}} & = & 
\langle r_1 \rangle
                        \left. \frac{d \ln T_1^r}{dN}
                        \right|_{N = \tilde{N}}
                        + 2 \nu_w(E_R) T_1^r \eta_l (1 - \tilde{f}_l)
\label{taur1dim}
.\end{eqnarray}

If space charge effects are not relevant, the first righthand terms of
both (\ref{taug1dim}) and (\ref{taur1dim}) vanish, so that for $\tau_1$
we have
\begin{equation}
\frac{1}{\tau_1} = \frac{1}{\tau_{g_1}} +
\frac{1}{\tau_{r_1}} = 2 \nu_w(E_R) \eta_l 
[ T_1^g \tilde{f}_l + T_1^r(1-\tilde{f}_l)]
.\label{tau1dim}
\end{equation}
The same passages can be repeated for the second barrier, once we
simply substitute all pedices $l$ with $r$ and $1$ with $2$.
Since we are considering the case of high bias, where
$\tilde{f}_r = 0$ and 
$\eta_r = 1$, we have $\langle r_2 \rangle = 2 \nu_w(E_R) T_2^r N$
and $1/\tau_2 = 2 \nu_w T_2^r$. By substituting these results, 
(\ref{r1dim_short}), and (\ref{tau1dim}) into (\ref{gamma}), we
obtain
\begin{equation}
\gamma  =  1 - \frac{2 T_2^r T_1^g \eta_l \tilde{f}_l}
	{\{T_2^r + \eta_l [ T_1^g \tilde{f}_l + T_1^r (1 - \tilde{f}_l)]\}^2}
\label{gammarecovery1}
.\end{equation}
It is straightforward to see that, because of the
term $\eta_l T_1^r (1-f_l)$ added to the denominator
of (\ref{gammarecovery1}), $\gamma$ cannot be smaller
than 1/2 (a closer look at (\ref{gammarecovery1}) could also
prove that $\gamma$ cannot be smaller than $1-\tilde{f}_l/2$).

Eq. (\ref{gammarecovery1}) becomes more readable if we consider particular
cases. First, let us suppose
that all available states in the well are above the conduction
band edge of the left electrode so that we have $\eta_l = 1$. 
This case corresponds to applied voltages V
smaller than the first peak voltage of the I--V characteristic.  
If, in addition, $\rho'_w$ has a narrow peak for $E=E_R$, we also 
have $T_1^g = T_1^r = T_1(E_R)$, and $T_2^r = T_2(E_R)$, that,
after substitution in (\ref{gammarecovery1}), yields
\begin{equation}
\gamma = 1 - \frac{2 T_1(E_R) T_2(E_R) \tilde{f}_l}
{(T_2(E_R) + T_1(E_R))^2}
= 1 - \tilde{f}_l \frac{T^{pk}_{cohe}}{2}
\label{gammarecovery2}
,\end{equation}
where $T^{pk}_{cohe}$ is the peak tunneling probability of the
double barrier in the case of completely coherent transport,
i.e., $T^{pk}_{cohe} = 4T_1T_2/(T_1+T_2)^2$.
However, the presence of $T^{pk}_{cohe}$ does not
mean that complete coherence is required for the applicability
of (\ref{gammarecovery2}): it is simply required that the
longitudinal density of states in the well has a resonance
narrow enough that 
all quantities involved in the calculation of transition rates
are practically constant in the energy range of the resonant peak.

Eq. (\ref{gammarecovery2}) shows the way the product 
$\tilde{f}_l T_{\rm cohe}^{\rm pk}$ affects  
the value of $\gamma$: maximum suppression
($\gamma = 1/2$) appears when both $\tilde{f}_l$ 
and $T_{\rm cohe}^{\rm pk}$ are 1, i.e., when
there is large charge accumulation at the emitter 
(high $\tilde{f}_l$), and symmetric barrier
transmission probabilities ($T_{\rm cohe}^{\rm pk} \approx 1$).
The devices characterized in Ref. \onlinecite{ciambrone95}
were designed to meet these conditions near the current
peak, where they actually exhibit a noise factor $\gamma \approx 1/2$.

If the $\Omega_l$-states that can be transmitted are completely filled,
we have $\tilde{f}_l = 1$, and we recover the well known result
\cite{li90,yurkkoch90,buettiker90,chenting91,buettiker92,davihyld92,liu95}
\begin{equation}
\gamma = 1 - \frac{T^{pk}_{cohe}}{2}
.\label{gammarecovery3}
\end{equation}

\subsubsection{Temperature dependence}

Recent experimental measurements of shot noise in double barrier diodes
as a function of temperature\cite{ciambrone95} have shown 
reduced suppression with increasing temperature. 
This fact has received little theoretical attention. A simple
explanation of this effect is provided by our model:
as temperature increases, inelastic collisions in the well 
become more frequent; in other words, the effective mean free
path gets shorter, resulting in  a lower and wider  
resonant peak in the density of states and a lower 
peak-to-valley ratio in the I-V characteristics.

We now have to remove the 
hypothesis of narrow density of states $\rho'_w$, while keeping
$\eta_l = 1$. Let us suppose
to be in the condition of high accumulation at the emitter, in other
words, let $\tilde{f}_l = 1$. From 
(\ref{gammarecovery1}) we obtain
\begin{equation}
\gamma = 1 - \frac{2 T_2^r T_1^g}{(T_2^r + T_1^g)^2}
\label{gammarecovery4}
.\end{equation}

In order to better explain the meaning of (\ref{gammarecovery4}),
let us consider the following situation: a constant voltage 
is applied between the electrodes, and the temperature is
progressively raised.
Suppose that the barrier dimensions are such that 
$T_1(E_R) \approx T_2(E_R)$ (this is the 
case, for example, of the devices fabricated and studied
by Ciambrone {\em et al.} \cite{ciambrone95}).
For very low
temperatures the hypothesis of narrow density of states is 
valid, because inelastic collisions are rare,
therefore Eq. (\ref{gammarecovery3}) is applicable,   
and a suppression factor of one half is expected.
When the temperature increases, the peak of $\rho'_w$ widens, and $T_1^g$ and
$T_2^r$ start to differ from $T_1(E_R)=T_2(E_R)$. In particular,
we would have $T_1^g > T_1(E_R) \approx T_2(E_R) > T_2^r$, 
because collisions
with phonons make electron in the well
relax to lower energy states, so that generation occurs more
easily at higher energies (because higher energy states in the well
are depopulated), while recombination occurs at lower energies
(because electrons occupy lower energy states).
It is straightforward to see from (\ref{gammarecovery4}) that 
$\gamma$ depends only on the ratio $T_2^r/T_1^g$, and is minimum
when that ratio is unity. 
As temperature increases, this ratio decreases, and $\gamma$
approaches unity. 
  
This interpretation is supported by the experimental
results shown in Ref. \onlinecite{ciambrone95}, where,
at temperatures up to 155~K, shot noise suppression
is smoothly dependent on temperature, while it rapidly 
vanishes at higher temperatures.

We would like to emphasize that, if $T_1(E_R) < T_2(E_R)$, 
it is possible to observe reduced shot noise with increasing 
temperature: as a matter of fact, since $T_1^g > T_1(E_R)$ 
and $T_2(E_R) > T_2^r$,
at some temperature we could have $T_1^g \approx T_2^r$, and
have a value of $\gamma$ close to one half.
Let us point out that the possibility of both positive
and negative dependence of shot noise on temperature
is one of the relevant prediction of our approach.

We also want to stress the point that shot noise cannot
be used as a probe for measuring the coherence of transport
in resonant tunneling diodes. In fact, 
in the so-called ``sequential regime''\cite{ianna_unified95,buettiker88}
all electrons entering the well are inelastically
scattered, but the density of states in the well is still
narrow enough that the shot noise
factor $\gamma$ is not affected by the collision rates and
is simply given by (\ref{gammarecovery2}), as in the 
case of completely coherent transport.

\subsubsection{Second resonance}

Very small noise reduction is to be expected after the first current peak:
electrons in the well relax towards the resonant peak
of $\rho'_w$,
and therefore fall below the conduction band edge of $\Omega_l$, leaving 
higher states in the well mostly empty, so that Pauli exclusion is poorly
effective in preventing generation from barrier 1. 
Many states in the well now have longitudinal energies
smaller then $E_{cbl}$, so that we have $\eta_l$ appreciably
smaller than unity. Let us also consider the case of high
accumulation, i.e., $\tilde{f}_l = 1$:  
(\ref{gammarecovery1}) becomes
\begin{equation}
\gamma = 1 - \frac{2 T_2^r T_1^g \eta_l}{(T_2^r + T_1^g \eta_l)^2}
\label{gammarecovery5}
,\end{equation}

What happens now is that maximum suppression can only be obtained
if $T_2^r = T_1^g \eta_l$. If energy relaxation cannot be disregarded,
most electrons in the well relax towards the first resonant peak
of $\rho'_w$,
and therefore fall below the conduction band edge of $\Omega_l$, leaving
higher states in the well empty, which means $\eta_l \approx 0$. In
this case, unless barrier transmission probabilities differ by
many orders of magnitude, no noise suppression should be obtained.
In other words, Pauli exclusion is no more effective in preventing
generation from barrier 1. This is the case, for example of the diodes studied by 
Ciambrone et al. \cite{ciambrone95}
 
As we have said in the introduction, noise characterization
can provide information about transport in such devices complementary
to that given by DC and AC characteristics. For example, 
if the density of states in the well $\rho'_w$ has a second
resonant level, shot noise measurements can provide useful insights
into the coherence of transport. In fact, in the case of
completely coherent transport, electrons do not relax to the
lower resonant peak of $\rho'_w$, and $\eta_l$ is close to one,
leading to a one half shot noise factor, if the barriers have equal 
transmission probability at the second resonant energy. Otherwise,
even a small rate of incoherent processes is sufficient to relax
electrons to lower levels, depopulating the second resonant level and
leading to an $\eta_l$ much smaller than 1 and to a shot noise factor close
to 1. 

Moreover, looking at the dependence of shot noise on temperature,
we could also determine whether Pauli exclusion or Coulomb repulsion
is the dominant cause of correlation
between different pulses: in fact, while the effectiveness
of Pauli exclusion is strongly dependent on the collision rate in
the well (i.e., on temperature) Coulomb repulsion 
(that is not accounted for in (\ref{gammarecovery1})) 
essentially depends only on the total
charge accumulated in the well. 

\subsection{Short time of observation}

Suppose that the time of observation $T$ is much smaller than the
characteristic times for generation and recombination through either
barriers, i.e., $\tau \gg T$. In this situation,
consecutive sub-pulses corresponding to 
a single electron traversing the device are separated by a time
longer than $T$ and do not appear to be correlated. This is easily the case, for
example, if the well region has macroscopic dimensions, i.e., 
several electron diffusion lengths; in such a case,
noise corresponding to two single barrier diodes in series is expected.

Let us recall that (\ref{hrromega}-\ref{hglrmomega}) are obtained for
$\omega \tau \ll 1$, which is not applicable now. Rather, we are
in the opposite limit of $\omega \tau \gg 1$, where
$h_{\alpha \beta}(t) = \langle \alpha \rangle$, ($\alpha,\beta=
r_1,r_2,g_1,g_2$), and $H_{\alpha \beta}(\omega) = 2\pi\delta(\omega)\langle
\alpha \rangle$,
an intuitive result that can be rigorously obtained from (\ref{hrlrm3}).
By substituting such form of $H_{\alpha \beta}$ in (\ref{c1})
and then in (\ref{sdiomega}) we have
\begin{equation}
\frac{S(\omega)}{2 q^2} = 
\lambda_1^2 \langle g_1 + r_1 \rangle +
\lambda_2^2 \langle g_2 + r_2 \rangle
\label{slong}
.\end{equation}
The full shot noise for the single barrier $m$ ($m=1,2$) in the
case of uncorrelated pulses is
given by $S_m(\omega) \equiv 2q (q\langle g_m \rangle +
q\langle r_m \rangle)$, therefore (\ref{slong}) can be written as
\begin{equation}
S(\omega) = \lambda_1^2 S_1(\omega) + \lambda_2^2 S_2(\omega)
\label{slong2}
.\end{equation}
Let us recall that $\lambda_m$ is the ratio of the voltage drop accross
barrier $m$ to the voltage drop across the whole device, (i.e., if
we refer to Fig. 3, $\lambda_m = \epsilon_m/qV$). 
Near equilibrium, if $R_m$ is the differential resistance of barrier $m$,
and $R = R_1 +R_2$ is the total device resistance,
we have $\lambda_m = R_m/R$ and $S_m = 4 k_B \Theta / R_m$
(as can be obtained from (\ref{utile})), which, substituted
in (\ref{slong2}), yields the Johnson-Nyquist result for $S(\omega)$,
as expected. 

If the bias is high enough that tranport occurs in only one direction,
i.e., for example, $\langle r_1 \rangle = \langle g_2 \rangle = 0$,
we have $S_1 = S_2 = 2 q \langle i \rangle$, with
$\langle i \rangle = q \langle g_1 \rangle = q \langle r_2 \rangle$
and 
\begin{equation}
\gamma = \frac{S(\omega)}{2 q \langle i \rangle} = 
\lambda_1^2 + \lambda_2^2 \geq \frac{1}{2}
,\end{equation}
where the equal sign holds if $\lambda_1 =\lambda_2 = 1/2$, i.e.,
again, if the structure is symmetric.

\section{Summary}

We have shown that our approach has a definite
advantage over those existing 
in the literature, because it can be applied in the whole
range of transport regimes. 
We have seen that the results of Ref.
\onlinecite{yurkkoch90,buettiker90,chenting91,buettiker92,davihyld92,liu95},
which are valid in the case of completely coherent transport,
are recovered, as well as those of Davies {\em et al.},\cite{davihyld92} which have been
obtained on the basis of a semiclassical model, i.e., for a high rate
of incoherent processes in the well. Our model has enabled us to
explain how similar results could be obtained from quite different models,
and even if correlation between consecutive sub-pulses was discarded.

Moreover, we have included in our model the combined effects
of Pauli exclusion and of Coulomb repulsion on the suppression
of shot noise, and concluded that in
practical devices a suppression in excess of one
half is not to be expected.

We have also studied shot noise in different operating conditions.
At equilibrium we have recovered the Johnson-Nyquist noise, which is
not a new result, but a good test for the validity of our model.
At sub-peak voltages we have predicted smaller suppression of shot
noise, due to the empty states at the cathode. At voltages
higher than that for the first peak, except for particular cases
implying strongly asymmetric barriers, we recover full shot noise,
because time correlations between transitions into and from the well
region are reduced due to electron thermalization. In particular,
for second resonance biases, we have shown that 
the Pauli exclusion plays no role in reducing shot noise,
if collisions are effective in establishing a thermal equilibrium
energy distribution; in that case,
the study of noise suppression in that case helps
us to determine the role played by Coulomb repulsion.

Dependence of shot noise suppression on temperature
has been observed in experiments,
and has been addressed theoretically for the first time in this paper. We have
shown that our model simply explains the experimental results.

In the future, numerical simulations of realistic resonant tunneling
structures based on our model will be performed, in order to
relax some of the approximations required to treat the problem analytically
and to make a comparison with experimental results.
In addition, we plan to perform 
experiments based on some new predictions of our model:
in particular, double barrier diodes with
barriers of equal transparency at the second resonance voltage peak,
and strongly asymmetric diodes which could exhibit enhanced shot noise
suppression with increasing temperature.

\section{Acknowledgments}

This work has been supported by the Ministry for the University
and Scientific and Technological Research and by the Italian
National Research Council (CNR) of Italy.
            
\appendix

\section{Correlation functions}

In this appendix we shall derive the
correlation functions $h_{\alpha \beta}(t)$ needed in Section III,
following Davies {\em et al.}\cite{davihyld92}.
Let us start by calculating
the correlation function $h_{r_l r_m}(t)$,
($l,m = 1,2$).
We need to define the conditional probability $p(N,t|M,0)$ that 
$N$ electrons are in the well at time $t$, given that
there were $M$ electrons at time 0. 
From it, we can write the conditional probability per unit time
$r_l(t|M,0)$ that an $r_l$-transition occurs at time $t$,
given that $M$ electrons were in the well at time $0$, 
in the form
\begin{equation}
r_l(t|M,0) = \sum_{N=0}^{N_0}
		r_l(N) p(N,t|M,0)
\label{rltm0}
,\end{equation}
where $N_0$ is the total number of states in the well.

We already introduced the
probability $p_0(M)$ of having $M$ electrons in the well, therefore
we can write
\begin{equation}
\langle r_m \rangle = \sum_{M=0}^{N_0} r_m(M) p_0(M)
\label{rmavg}
,\end{equation}
from which we obtain $p_{r_m}(M)$, i.e., 
the probability that, when an $r_m$-hop occurs, $M$ electrons
are left in the well, as
\begin{equation}
p_{r_m}(M) \equiv
\frac{p_0(M+1) r_m(M+1)}{\langle r_m \rangle}
\label{prm}
\end{equation}
The probability per unit time that 
an $r_l$-hop occurs at time $t$, given that an $r_m$-hop occurred at 
time $0$, i.e., $h_{r_l r_m}(t)$, is then given by
\begin{equation}
h_{r_l r_m}(t) =
\sum_{M=0}^{N_0}
p_{r_m}(M) r_l(t|M,0)
\label{hrlrm}
\end{equation}

From (\ref{rlin}) we can write (\ref{rltm0}) as
\begin{eqnarray}
r_l(t|M,0) & = & \sum_{N=0}^{N_0}
	\left( \langle r_l \rangle + \frac{\Delta N}{\tau_{rl}} \right)
	p(N,t|M,0) \nonumber \\
& = & \langle r_l \rangle
	 + \frac{M-\langle N \rangle}{\tau_{rl}} e^{-|t|/\tau}
\label{rltm0bis}
,\end{eqnarray}
where the second equality comes from the fact that $p(N,t|M,0)$
is normalized to unity and that the number of electrons in the
well relax exponentially with time constant $\tau$ to the
mean value $\langle N \rangle$,\cite{davihyld92}
as we know from the rate equation $dN/dt = g(N) -r(N)$.

From (\ref{rlin}), (\ref{rmavg}-\ref{rltm0bis}), we now write
\begin{eqnarray}
h_{r_l r_m}(t) & = & \frac{1}{\langle r_m \rangle}
			\sum_{M=0}^{N_0} p_0(M+1) \times
			\nonumber \\
& &\left(\langle r_m \rangle + \frac{M+1-\langle N \rangle}{\tau_{rm}}
	\right)
	\left( \langle r_l \rangle
		 + \frac{M-\langle N \rangle}{\tau_{rl}} e^{-|t|/\tau}
	\right) \nonumber \\
& = & 
\langle r_l \rangle - \frac{1}{\tau_{rl}}  
\left( 1 - \frac{\mbox{var}(N)}{\tau_{rm} \langle r_m \rangle} \right) 
e^{-|t|/\tau}
\label{hrlrm3}
;\end{eqnarray}
sums are easily evaluated if we notice that 
$\sum_{N=0}^{N_0} p_0(N) \Delta N = 0$ and
$\sum_{N=0}^{N_0} p_0(N) (\Delta N)^2 = \mbox{var}(N)$.
For $\omega \ll 1/\tau$, the 
Fourier transform $H_{r_l r_m}(\omega)$ of (\ref{hrlrm3}) is
\begin{equation}
H_{r_l r_m}(\omega) = 
2 \pi \delta(\omega) \langle r_l \rangle - \frac{2\tau}{\tau_{rl}}
\left( 1 -\frac{\mbox{var}(N)}{\tau_{rm} \langle r_m \rangle} \right)
\label{hrromega}
.\end{equation}

Analogously, we can obtain the correlation functions for all the other
processes, i.e., for $l,m = 1,2$:
\begin{eqnarray}
H_{g_l g_m}(\omega) & = & 2 \pi \delta(\omega) \langle g_l \rangle
- \frac{2 \tau}{\tau_{gl}}
\left( 1 - \frac{\mbox{var}(N)}{\tau_{gm} \langle g_m \rangle} \right)\\
H_{r_l g_m}(\omega) & = & 2 \pi \delta(\omega) \langle r_l \rangle
+ \frac{2 \tau}{\tau_{rl}}
\left( 1 - \frac{\mbox{var}(N)}{\tau_{gm} \langle g_m \rangle} \right)\\
H_{g_l r_m}(\omega) & = & 2 \pi \delta(\omega) \langle g_l \rangle
+ \frac{2 \tau}{\tau_{gl}}
\left( 1 - \frac{\mbox{var}(N)}{\tau_{rm} \langle r_m \rangle} \right)
\label{hglrmomega}
\end{eqnarray}

\section{Useful formulas}

\subsection{Calculation of $S(\omega)$}

From (\ref{iomegacompleta}),
by using (\ref{sumterm2}) and considering the fact that 
$\langle \beta \rangle H_{\alpha \beta}
= \langle \alpha \rangle H_{\beta \alpha}$ ($\alpha,\beta = g_1,g_2,r_1,r_2$),
we have 
\begin{eqnarray}
\frac{ \langle |I(\omega)|^2 \rangle }{q^2 T} & = & 
\lambda_1^2 \left\{ 
\langle g_1 \rangle [ 1 + H_{g_1,g_1}(\omega) ] - 2 \langle g_1 \rangle H_{r_1,g_1}(\omega) 
+ \langle r_1 \rangle [ 1 + H_{r_1,r_1}(\omega) ]
\right\} \nonumber \\
& + & 2 \lambda_1 \lambda_2 \left[
- \langle g_1 \rangle H_{g_2,g_1}(\omega)
+ \langle g_1 \rangle H_{r_2,g_1}(\omega)
+ \langle r_1 \rangle H_{g_2,r_1}(\omega)
- \langle r_1 \rangle H_{r_2,r_1}(\omega) \right] \nonumber \\
& + & \lambda_2^2 \left\{ 
\langle g_2 \rangle [ 1 + H_{g_2,g_2}(\omega) ] - 2 \langle g_2 \rangle H_{r_2,g_2}(\omega)
+ \langle r_2 \rangle [1 + H_{r_2,r_2}(\omega)] \right\}
\label{c1}
\end{eqnarray} 

Substitution of (\ref{hrromega}-\ref{hglrmomega}) in (\ref{c1}) and
then in (\ref{sdiomega}) yields
\begin{eqnarray}
\frac{S(\omega)}{2q^2} & = &
\lambda_1^2
\left[ 
\langle g_1 + r_1 \rangle \left( 1 - \frac{2\tau}{\tau_1} \right)
+ \frac{2 \tau}{\tau_1^2} \mbox{var}(N) 
\right] \nonumber \\
& & + 2 \lambda_1 \lambda_2 
\left( \frac{\tau}{\tau_1} \langle g_2 + r_2 \rangle
	+ \frac{\tau}{\tau_2} \langle g_1 + r_1 \rangle 
	- \mbox{var}(N) \frac{2\tau}{\tau_1 \tau_2} \right) 
		\nonumber \\
& & + \lambda_2^2 
\left[ \langle g_2 + r_2 \rangle \left( 1 - \frac{2\tau}{\tau_2} \right)
+ \frac{2\tau}{\tau^2} \mbox{var}(N) \right]
\label{c2}
;\end{eqnarray}
But, if we put (\ref{varn}) in (\ref{c2}), by writing $\langle g \rangle$
as $\langle g_1 + r_1 + g_2 + r_2 \rangle/2$, we obtain
\begin{equation}
\frac{S(\omega)}{2q^2} = (\lambda_1^2 + 2 \lambda_1 \lambda_2 +
\lambda_2^2 ) \left( \frac{\tau^2 \langle g_1 +r_1 \rangle}{\tau_2^2}
+ \frac{\tau^2 \langle g_2 + r_2 \rangle}{\tau_1^2} \right)
\label{c3}
,\end{equation}
which reduces to (\ref{sdiomegabella}) if we simply remember that 
$\lambda_1 + \lambda_2 = 1$. Note that, as expected, dependence 
of noise at low frequencies upon the relative sizes of current
pulses due to traversal of the two barriers, is cancelled out.

Now, from (\ref{il_ir}) and the fact that $r = r_1 + r_2$, 
we can write 
\begin{eqnarray}
q \langle g_1 + r_1 \rangle & = &
\left( \frac{2\langle r_1 \rangle}{\langle r_2 \rangle} + 1 \right)
I_l + I_r \nonumber \\
q \langle g_2 + r_2 \rangle & = &
\left( \frac{2\langle r_2 \rangle}{\langle r_1 \rangle} + 1 \right)
I_r + I_l
\label{g1piur1}
;\end{eqnarray}

By substituting (\ref{g1piur1}) and (\ref{tau}) 
in (\ref{sdiomegabella}), we straightforwardly get Eq.
(\ref{sdiomegailir}).
 
\subsection{Derivation of Eq. 45}

Each state $|\alpha_s \rangle$ ($s=l,w,r$)
is characterized by its total energy $E^T_s$ and a set of other
parameters $\beta_s$. Electron transitions
between regions conserve energy, hence we can
write $|M_{1lw}|^2 = |M_1(E^T_w,\beta_w,\beta_l)|^2 \delta(E_w-E_l)$.
From (\ref{g1iniz}) and (\ref{r1iniz}) we can write
\begin{eqnarray}
g_1 -r_1 & = & \frac{2\pi}{\hbar}
\int \int \int
|M_1(E^T,\beta_w,\beta_l)|^2 \rho_l(E^T,\beta_l) \rho_w(E^T,\beta_w) \times
						\nonumber \\
& &	[ f_{l}(E^T,\beta_l) - f_{w}(E^T,\beta_w) ] 
	d\beta_w d\beta_l dE^T
\label{g1menor1}
.\end{eqnarray}
Differentiation of (\ref{g1menor1})
with respect to $\epsilon_1 = E_{fl} - E_{fw}$,
given that at equilibrium $f_{l} = f_{w}$, yields
\begin{equation}
\left. \frac{d(g_1 -r_1)}{d\epsilon_1}
\right|_{\epsilon_1 =0} = \frac{2\pi}{\hbar}
\int \int \int
|M_1(E^T,\beta_w,\beta_l)|^2 \rho_l(E^T,\beta_l) \rho_w(E^T,\beta_w) 
\left. \frac{df_w}{dE_{fw}} \right|_{\epsilon_1 = 0}
d\beta_w d\beta_l dE^T
\label{dg1menor1}
,\end{equation}
in which, according to the Fermi-Dirac statistics which
holds at thermal equilibrium, we have 
\begin{equation}
\left. \frac{df_{w}}{dE_{fw}} \right|_{\epsilon_1 = 0}
= \frac{f_{w0}(1 - f_{w0})}{k_B \Theta}
\label{dfland}
,\end{equation}
where $k_B$, $\Theta$, and $f_{w0}$ are the 
Boltzmann constant, the temperature, and Fermi-Dirac occupation
factor, respectively. 
Substitution of (\ref{dfland}) in (\ref{dg1menor1}) finally yields
\begin{equation}
\left. \frac{d(g_1 -r_1)}{d\epsilon_1}
\right|_{\epsilon_1 =0} = 
\frac{g_1}{k_B \Theta}
\label{utile}
.\end{equation}

\begin{figure}
\vspace{2cm}

\epsfxsize=\figwidth
\epsffile{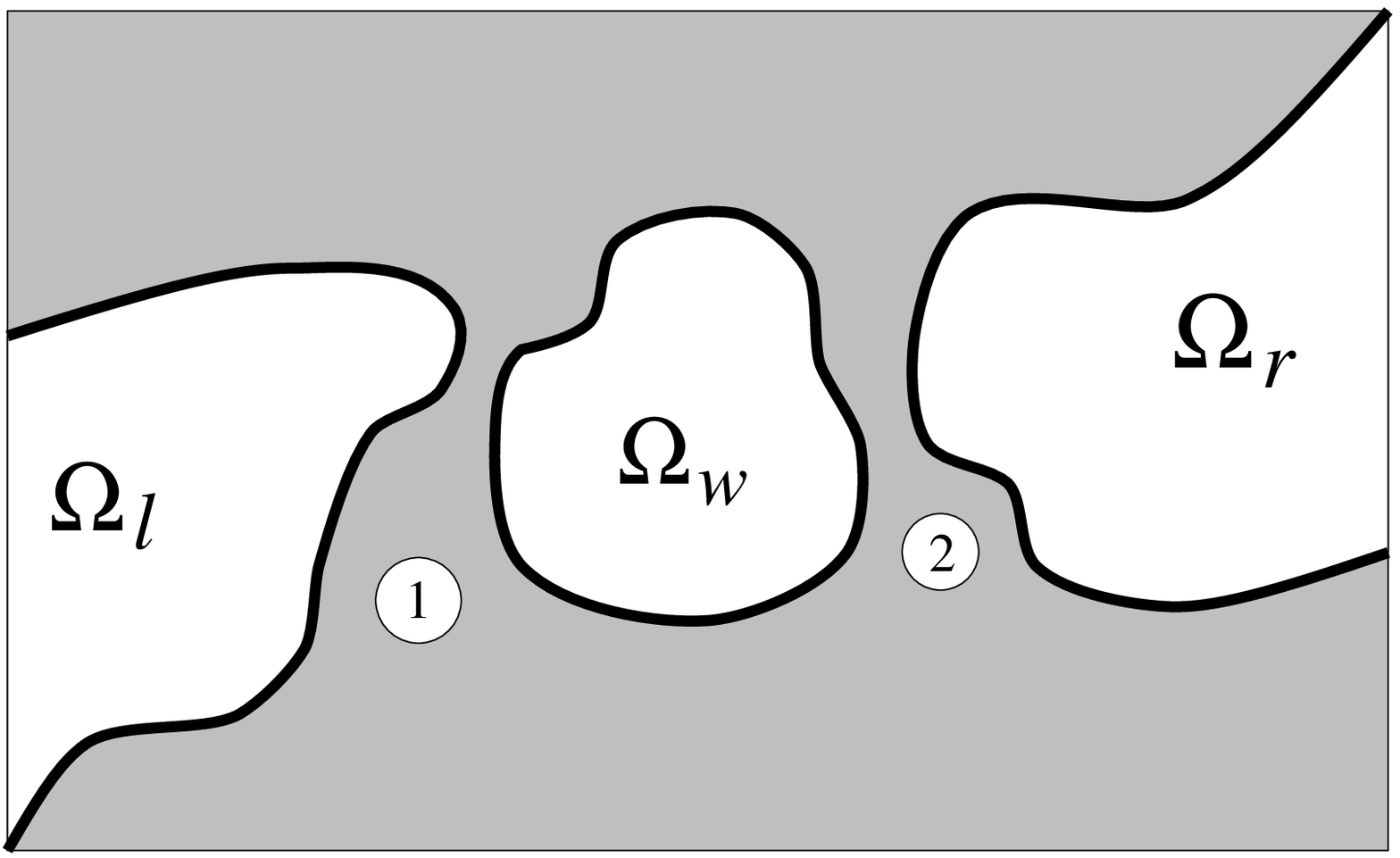}

\vspace{1cm}

\caption{A generic resonant tunneling structure consists of three
isolated regions $\Omega_l$, $\Omega_w$, $\Omega_l$ weakly coupled
by tunneling barriers, here indicated with 1 and 2. Coupling
between different regions has to be small enough to be treated with
first order perturbation theory.}
\end{figure}
\newpage
\begin{figure}

\epsfxsize=\figwidth
\epsffile{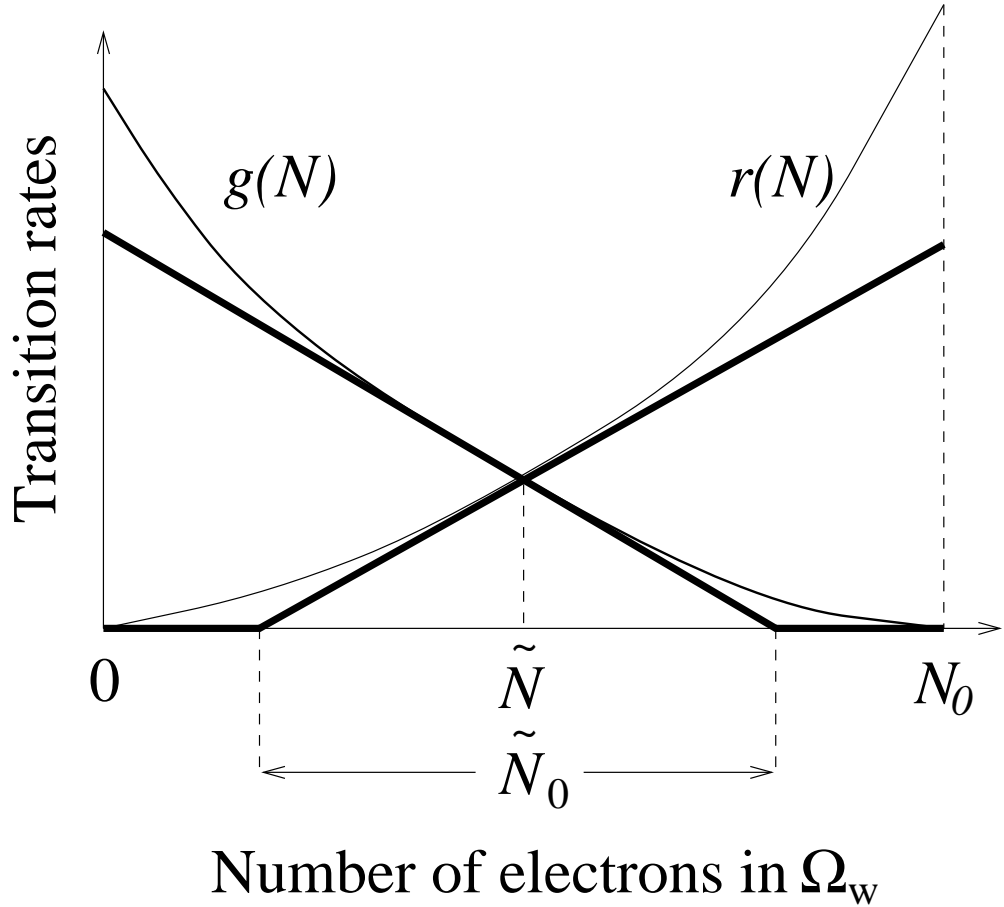}

\vspace{1cm}

\caption{A qualitative sketch of the total generation and recombination
rates is shown (thin lines), along with the linearization 
described in Sec. IID. $N_0$ is the total number
of states in the well, $\tilde{N}$ is the number of electrons for
which $g(\tilde{N}) = r(\tilde{N})$, $\tilde{N}_0$ is the maximum
allowed excursion for $N$ in the linearized approximation.}
\end{figure} 

\newpage
\begin{figure}
\epsfxsize=\figwidth
\epsffile{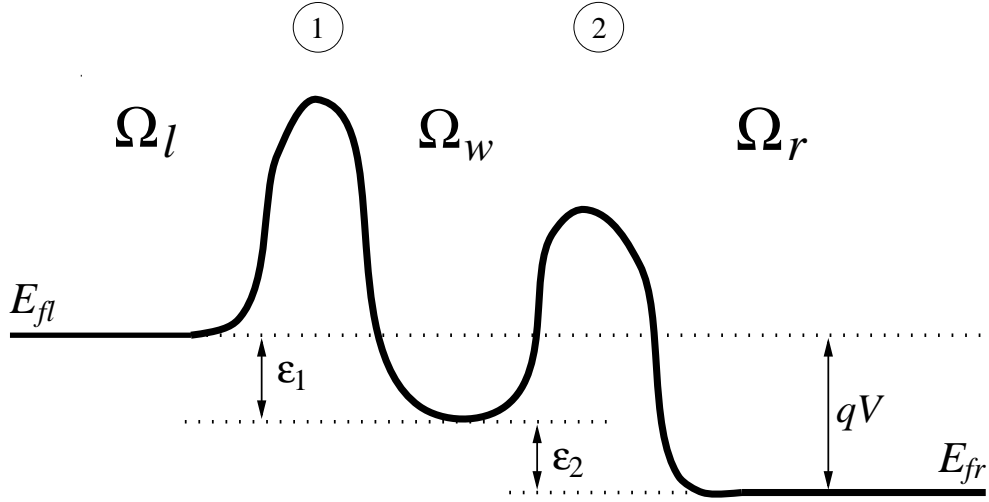}

\vspace{1cm}

\caption{The longitudinal section of a resonant tunneling
device is sketched; $E_{fl}$, $E_{fw}$, and $E_{fr}$ are the quasi-Fermi levels
of $\Omega_l$, $\Omega_w$, $\Omega_r$, respectively. A small voltage
$V$ is applied between the electrodes, and $\epsilon_1$, $\epsilon_2$
are the the partial potential energy drops across barriers 1 and 2, 
respectively}
\end{figure}

\end{document}